\documentclass[pra,onecolumn,floatfix,notitlepage,groupedaddress]{revtex4-1}

\usepackage[hidelinks,colorlinks=false,citecolor=black,linkcolor=black]{hyperref}
\usepackage{mathtools}
\usepackage{subcaption}
\usepackage{amssymb,amsmath,amstext,amsthm}
\usepackage{graphicx}
\usepackage{epstopdf}
\usepackage{color}
\usepackage{bm}
\usepackage{appendix}
\usepackage[T1]{fontenc}
\usepackage{bbm}
\usepackage{latexsym}

\usepackage{algorithm,algorithmic}

\usepackage{float}
\usepackage{verbatim}
\usepackage{lipsum}
\usepackage{braket}
\usepackage{ulem}

\newcommand{\del}[0]{\partial}

\newtheorem{lemma}{Lemma}

\begin{document}

\title{Saturating the one-axis twisting quantum Cram\'{e}r-Rao bound with a total spin readout}
\author{T.J.\,Volkoff}
\affiliation{Theoretical Division, Los Alamos National Laboratory, Los Alamos, NM, 87545 USA.}
\author{Michael J. Martin}
\affiliation{Materials Physics and Applications Division, Los Alamos National Laboratory, Los Alamos, NM 87545, USA.}

\begin{abstract}
We show that the lowest quantum Cram\'{e}r-Rao bound achievable in interferometry with a one-axis twisted spin coherent state is saturated by the asymptotic method of moments error of a protocol that uses one call to the one-axis twisting, one call to time-reversed one-axis twisting, and a final total spin measurement (i.e., a twist-untwist protocol). The result is derived by first showing that the metrological phase diagram for one-axis twisting is asymptotically characterized by a single quantum Fisher information value $N(N+1)/2$ for all times, then constructing a twist-untwist protocol having a method of moments error that saturates this value. The case of finite-range one-axis twisting is similarly analyzed, and a simple functional form for the metrological phase diagram is found in both the short-range and long-range interaction regimes. Numerical evidence suggests that the finite-range analogues of twist-untwist protocols can exhibit a method of moments error that asymptotically saturates the lowest quantum Cram\'{e}r-Rao bound achievable in interferometry with finite-range one-axis twisted spin coherent states for all interaction times.
\end{abstract}
\maketitle

\section{Introduction}
\label{sec:intro}
Since the first proposals for light-pulse atom interferometry for acceleration sensing \cite{PhysRevLett.67.181,Kasevich1992}, improvements to quantum sensing schemes with atomic ensembles have been obtained due to advances in controllability and readout \cite{RevModPhys.90.035005}. A major advancement in the controllability of these ensembles comes from methods to generate atom-atom entanglement in the ensembles, in particular by spin squeezing, in which the total spin fluctuations in one direction are suppressed while the spin fluctuations in an orthogonal direction are amplified \cite{PhysRevA.47.5138,PhysRevLett.122.223203,PhysRevA.96.050301}. Spin squeezing in ultracold gaseous atomic ensembles has led to $O(10)$ dB improvement over non-entangled counterpart for sensing small rotations \cite{Hosten2016} and $O(1)$ dB improvement in Allan deviation (i.e., frequency stability) of optical lattice clocks \cite{pp2020,Bao2020}. Spin squeezing of lattice Rydberg atoms has obtained $O(1)$ dB improvement for frequency estimation beyond the corresponding non-entangled system \cite{Eckner2023}. We consider both classes of systems in the present work; the results of Section \ref{sec:gt} and \ref{sec:vvv} are relevant to the setting of ultracold gaseous atomic ensembles and the results of Section \ref{sec:foat} are relevant to the setting of lattice Rydberg atoms.

The connection between spin squeezing of atomic ensembles and enhanced performance of interferometry comes from the fact that spin squeezing can reduce the lower bound for the estimation error given by quantum Cram\'{e}r-Rao inequality (QCRI) \cite{RevModPhys.90.035005}
\begin{equation}
(\Delta \tilde{\phi})^{2}\ge {1\over \nu F(\ket{\psi_{\phi}})}
\end{equation}
where $\tilde{\phi}$ is a classical estimator of the interferometric phase shift $\phi$, $(\Delta \tilde{\phi})^{2}$ is its variance, $\nu$ is the number of measurements carried out, and $F$ is the quantum Fisher information (QFI) for the phase-shifted state $\ket{\psi_{\phi}}$ defined by 
\begin{equation}
F(\ket{\psi_{\phi}}):= 4\text{Var}_{\ket{\psi}}\vec{n}\cdot \vec{J}
\label{eqn:qcriqcri}
\end{equation}
when the phase shift operation is given by the SU(2) rotation $R_{\phi}=e^{-i\phi \vec{n}\cdot \vec{J}}$, $\Vert \vec{n}\Vert=1$.\footnote{In the present work, we will consider the one-shot setting, $\nu=1$.} Specifically, whereas a non-entangled atom ensemble has a maximum QFI $N$ (called the standard quantum limit (SQL)), an entangled atom ensemble can have QFI $O(N^{2})$ (Heisenberg scaling), with maximum possible value $N^{2}$ (Heisenberg limit). Due to the fact that the QCRI (\ref{eqn:qcriqcri}) can be saturated by an optimal quantum measurement \cite{bc}, the QCRI implies that interacting quantum spin systems have the potential to allow estimation of SU(2) rotations with quadratically lower error (in the system size) than is achievable with non-interacting (i.e., unentangled) quantum spin systems. However, the optimal quantum measurement is not necessarily readily implementable in practice. The goal of the present work is to show that a recently introduced protocol allows to achieve the maximal QFI for spin squeezed states by using only a measurement of the total spin in a certain direction.

For a given SU(2) rotation $R_{\phi}$ defining the quantum estimation problem, construction of a protocol that approximately saturates the QCRI subject to dynamical and kinematical constraints of the quantum spin system is a central part of quantum metrology research in atomic systems.  A generic noiseless protocol can be constructed as
\begin{equation}
    \ket{\psi} \rightarrow U\ket{\psi} \rightarrow R_{\phi}U\ket{\psi} \rightarrow VR_{\phi}U\ket{\psi} \rightarrow f(\phi)
    \label{eqn:sch}
\end{equation}
where the steps correspond to 1. pre-rotation evolution of the probe quantum system, 2. SU(2) rotation, 3. post-rotation evolution of the probe, 4. measurement of a linear or nonlinear signal $f(\phi)$ from which to extract the rotation estimate. The signal depends on $\ket{\psi}$, $U$, $V$, and the chosen quantum measurement. In (\ref{eqn:sch}), the restriction to pure probe states and unitary evolutions is due to the fact that noisy quantum states never provide the greatest QFI defining the lowest lower bound in the QCRI for the error of a rotation estimation protocol.

Protocols that take $U$ and $V$ to be one-axis twisting evolutions \cite{PhysRevA.47.5138} or finite-range one-axis twisting evolutions have proven central to high-performance quantum rotation estimation protocols \cite{PhysRevLett.116.053601,winey,vuletic,PhysRevLett.117.013001,Schulte2020ramsey}. There are two physical reasons for this: 1. one-axis twisting generates spin squeezing, 2. one-axis twisting naturally appears as the effective interaction in low-energy, two-mode atomic systems. In fact, for interaction time $\pi/2$, one-axis twisting generates a Greenberger-Horne-Zeilinger (GHZ) state with direction dependent on the parity of $N$. Such a GHZ state allows to achieve the global Heisenberg limit for sensing an SU(2) rotation (global Heisenberg limit refers to the QFI being equal to $N^{2}$ for all $\phi$). To see this, consider the GHZ probe state evolving under a differential phase shift according to \cite{PhysRevLett.96.010401}
\begin{equation}
    \ket{\psi_{\phi}}:= e^{-i\phi J_{z}}{\ket{N,0}+\ket{0,N}\over \sqrt{2}}
\end{equation} where the occupation basis $\lbrace \ket{N-k,k}\rbrace_{k=0}^{N}$ (i.e., Dicke basis) is defined by the occupation of single-particle states $\ket{0}$, $\ket{1}$. Defining $X$, $Y$, and $Z$ to be the Pauli matrices in this basis, one can use a measurement of a global parity operator such as $X^{\otimes N}$ to construct an estimate of $\phi$ that saturates the Heisenberg limit. Specifically, when $X^{\otimes N}$ is measured, the method of moments estimator $\tilde{\phi}$ \cite{RevModPhys.90.035005} of the rotation angle $\phi$ has asymptotically normal distribution with variance given by
\begin{eqnarray}
    (\Delta \tilde{\phi})^{2} &:=& {\text{Var}_{\ket{\psi_{\phi}}}X^{\otimes N}\over \left( \del_{\phi}\langle X^{\otimes n}\rangle_{\ket{\psi_{\phi}}} \right)^{2}} \nonumber \\
    &=& {1-\langle X^{\otimes N}\rangle_{\ket{\psi_{\phi}}}^{2}\over \left( \del_{\phi}\langle X^{\otimes N}\rangle_{\ket{\psi_{\phi}}} \right)^{2}} \nonumber \\
    &=& {1\over N^{2}}
    \label{eqn:kkl}
\end{eqnarray}
for all $\phi$. Note that the method of moments error is given by a formula that propagates the error in estimating the expectation of an observable to the error in the estimator of the parameter. Although the above example protocol has no faults in a digital quantum device that utilizes, e.g., a Hadamard and CNOT circuit to generate the GHZ state and carries out measurement of an $N$-qubit Pauli group observable, the example may be infeasible in an analog quantum system such as a gas of itinerant atoms, in which a linear depth circuit involving pairwise-entangling gates and measurements of the individual particles cannot be implemented due to a lack of addressability. One could note that the same method of moments sensitivity can be obtained by measuring the symmetric operator $\ket{N,0}\bra{0,N}+h.c.$ \cite{dowl}, which does not require a measurement to be carried out on individual atoms. Despite this fact, if the measurement statistics of this operator are to be obtained by counting the occurrences of $\ket{N,0}$ and $\ket{0,N}$ in a measurement of the difference of occupation number between the modes, a GHZ-generating circuit must still appear somewhere in the protocol to convert the eigenvectors of $\ket{N,0}\bra{0,N}+h.c.$ to those of $\ket{N,0}\bra{N,0}+\ket{0,N}\bra{0,N}$. 

The problem of generating the GHZ state in an analog system is usually solved by appealing to the fact that certain spin-squeezing entanglement generation processes can produce GHZ states at long interaction times. Specifically, by the Yurke-Stoler process in the spin setting \cite{PhysRevA.40.2417}, one finds that, for even $N$, one-axis twisting dynamics $e^{-itJ_{z}^{2}}$ generates a GHZ state in a given direction on the Bloch sphere from a spin coherent state in that direction for $t=\pi/2$. For odd $N$, the resulting GHZ state is rotated by $\pi / 2$ relative to the initial axis. The quality of the GHZ state depends on whether the interaction time $t=\pi /2$ can be reached with high state fidelity. However, the problem of finding an alternative measurement scheme remains. One class of methods to resolve this problem involves applying a second one-axis twisting operation to create the probe state $e^{it_{2}J_{z}^{2}}$, and replacing the global measurement $X^{\otimes N}$ by a measurement of a total spin operator $\vec{n}\cdot \vec{J}$ to yield an estimate of the probability distribution $p_{m}(\phi)$, $m=\lbrace-{N\over 2},-{N\over 2}+1,\ldots,{N\over 2}\rbrace$ \cite{PhysRevLett.119.193601}. In this case, the $\phi$ signal is the probability density corresponding to a total spin measurement. 

On the other hand, in practice it is desirable to design quantum metrology protocols that obtain a parameter estimate from the minimal possible information obtainable from a measurement, e.g., that use low moments of a total spin operator as the $\phi$ signal.
For instance, it has been noted that Heisenberg scaling can be attained by carrying out total spin measurement on a probe state prepared by two calls to one-axis twisting dynamics (with opposite interaction signs and small interaction times $t\sim {1\over \sqrt{N}}$) \cite{PhysRevLett.116.053601}. Specifically, such a probe state is prepared by twist-untwist protocol, which we define as
\begin{equation}\ket{\psi_{\phi}(\vec{n})}=e^{itJ_{z}^{2}}e^{-i\phi \vec{n}\cdot \vec{J}}e^{-itJ_{z}^{2}}\ket{\zeta =1}.\label{eqn:ps}\end{equation}
From the probe state (\ref{eqn:ps}) an estimate of $\phi$ is obtained by measurement of the total spin operator $\vec{m}\cdot\vec{J}$. In (\ref{eqn:ps}), we use the SU(2) coherent state
\begin{equation}
    \ket{\zeta}={1\over (1+\vert \zeta\vert^{2})^{N/2}}\sum_{\ell=0}^{N}\sqrt{N\choose \ell}\zeta^{\ell}\ket{N-\ell,\ell} 
\end{equation}
with $\zeta=1$, where $\ket{N-\ell,\ell}$ is the Dicke state (i.e., number occupation basis state) with $\ell$ particles in the atomic state $\ket{1}$. More generally,  $\zeta \in \mathbf{C}\cup \lbrace \infty\rbrace$ is the stereographic projection from the south pole of the Bloch sphere. Note that in (\ref{eqn:ps}), the dependence of the state on the particle number $N$ and interaction time $t$ has been suppressed in the notation of the right hand side. In (\ref{eqn:ps}), it is natural to ask why the untwist operation is taken to have the same interaction time $t$ as the twist operation. For the smallest interaction times which allow Heisenberg scaling (namely, $t=O(N^{-1/2})$), optimality of this convention has been shown \cite{PhysRevResearch.4.013236}. The possibility of  untwist layer of an optimal protocol being defined by a different interaction time compared to the initial twist layer was explored in Ref.\cite{Schulte2020ramsey}, including in the presence of collective dephasing noise. In the present work we show analytically that the method of moments error for an optimal twist-untwist protocol (with untwist interaction time the same as the twist interaction time) saturates the lowest possible QCRI for the one-axis twisting path $e^{-itJ_{z}^{2}}\ket{\zeta =1}$ asymptotically in $N$, regardless of the interaction time. This result is obtained by carefully computing the $\phi \rightarrow 0$ limit of the method of moments error in a region of the metrological phase diagram that covers all interaction times as $N\rightarrow \infty$.  We provide numerical evidence for the same result when the one-axis twisting generator $J_{z}^{2}$ is replaced by a finite-range one-axis twisting generator with range $K$ spin interactions on a one-dimensional lattice \cite{PhysRevLett.112.103601}. These results show that twist-untwist protocols, which utilize a total spin readout but require control of the sign of atomic interactions, are metrologically equivalent to a protocol that performs optimal quantum estimation of the rotation angle of a one-axis twisted probe state about its most sensitive axis.

A brief outline of the results is as follows: Section \ref{sec:back} provides background on the asymptotic  (in the number of measurement shots) method of moments error that is used to quantify the sensitivity of twist-untwist protocols in this manuscript. We recall that a one-axis twisted probe state at interaction time $\pi/2$ is known to have a QCRI given by the Heisenberg limit $1/N^{2}$, and subsequently show that the method of moments error quantifier for a twist-untwist protocol (with the same interaction time) and a particular total spin measurement saturates this Heisenberg limit.  Section \ref{sec:gt} derives a metrological phase diagram for one-axis twisting building on previous work analyzing the asymptotics of QFI and spin squeezing in this system \cite{ps,PhysRevLett.122.090503,crphys,PhysRevLett.127.160501}. It is also shown that for large $N$, the metrological phase diagram degenerates into a QFI plateau with value $N(N+1)/2$, i.e., the QFI takes this value for all interaction times except a set of asymptotic measure zero. In Section \ref{sec:vvv} we show that combining a twist-untwist protocol with a $J_{x}$ measurement gives a method of moments error that asymptotically saturates the QCRI on this QFI plateau, thereby showing that twist untwist protocols with total spin readout are asymptotically sufficient for estimation of SU(2) rotations with one-axis twisted probe states for almost all interaction times.  Section \ref{sec:foat} derives a metrological phase diagram for finite-range one-axis twisting for both cases of short-range and long-range interactions, with subsection \ref{sec:frtwun} providing numerical evidence that a twist-untwist protocol and total spin measurement saturates the maximal QFI for a finite-range one-axis twisted probe state, at least for a large range of interaction times.

\begin{figure*}[t!]
\centering
\includegraphics[scale=0.7]{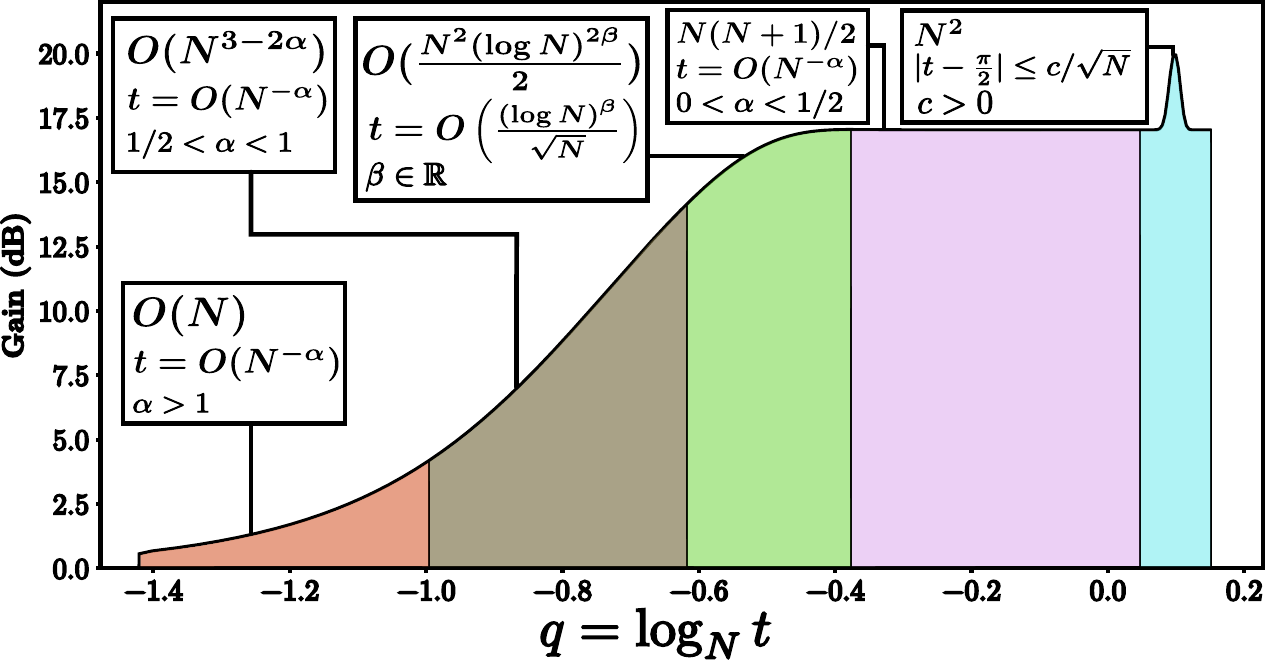}
    \caption{Metrological phase diagram for the one-axis twisted state $e^{-itJ_{z}^{2}}\ket{\zeta=1}$ with interaction time $t={N^{q}}$ (the black curve is the maximal QFI over $\vec{n}$ for $N=100$).  There exists an interpolating interaction time scaling for adjacent phases. In each box, the first line shows the QFI scaling, the second line shows the corresponding interaction time scaling, and the third line shows the domain of parameters appearing in the interaction time scaling. On a linear plot, the pink phase is nearly the whole time domain, as expected from Lemma \ref{lem:11}.
}
    \label{fig:schem}
\end{figure*}

\subsection{Relation to previous work}
To the knowledge of the authors, Ref.\cite{PhysRevA.92.043622} was the first work which optimized, for arbitrary interaction time $t$, the QFI for a one-axis twisted SU(2) coherent state probe \begin{equation} 
e^{-itJ_{z}^{2}}\ket{\zeta=1}\label{eqn:rere}\end{equation} over all possible unitary parametrizations of the form $e^{-i\phi\vec{n}\cdot \vec{J}}$, with $\vec{n}$ a unit three-vector. That same work also recognized the dominance of a single metrological phase for the optimal QFI for such probe states, but without providing a rigorous argument as in Lemma \ref{lem:11} below. An analytical extension of that work to all interaction times is the motivation for Section \ref{sec:gt}. The black curve in Fig. \ref{fig:schem} appears in Ref.\cite{PhysRevA.92.043622,PhysRevLett.122.090503}. Further, Ref. \cite{PhysRevLett.122.090503} (Ref.\cite{crphys}) provided numerical (analytical) evidence of the insufficiency, for interaction times $t\sim  O(N^{-\alpha})$, $\alpha < 2/3$, of method of moments estimation with a total spin observable for saturating the maximal QFI of probe state (\ref{eqn:rere}) over all directions $\vec{n}$.

For an atomic ensemble, Ref.\cite{PhysRevLett.116.053601} demonstrated Heisenberg scaling of the reciprocal of the method of moments error with a total $J_{y}$ spin readout after applying a twist-untwist protocol (\ref{eqn:ps}) with $\vec{n}=(0,1,0)$, although this error does not saturate the QCRI. Working with the same total spin $J_{y}$ readout as Ref.\cite{PhysRevLett.116.053601}, but allowing the final one-axis twisting layer to have a different interaction time compared to the first layer, Ref.\cite{PhysRevResearch.4.013236} established the following fact: if both interaction times scale as $O(N^{-1/2})$ (an interaction time domain in which  both the QFI and the reciprocal method of moments error with $J_{y}$ readout scale as $O(N^{2})$, although the latter does not saturate the former), then the final one-axis twisting layer should indeed be taken to have the same interaction time magnitude as the first layer, but with opposite interaction sign (i.e., a twist-untwist protocol is optimal for sensing a $J_{y}$ rotation with a $J_{y}$ readout in this interaction time domain). It was shown in Ref.\cite{PhysRevLett.127.160501} that the final untwisting interaction in (\ref{eqn:ps}) allows one to use method of moments estimation with a total spin observable to saturate, for various interaction times, an optimal spin-squeezing quantifier for the state (\ref{eqn:rere}). Section \ref{sec:vvv} of the present work is framed in the same setting of a twist-untwist protocol with a total spin readout, and as we discuss further in Section \ref{sec:back}, the methods of moments error must be calculated carefully in a neighborhood of $\phi=0$ in order to show that the reciprocal of the method of moments error saturates the optimal QFI of (\ref{eqn:rere}) over $\vec{n}$ for large $N$. Our results in Section \ref{sec:vvv} expand both the numerator and denominator of the method of moments error (\ref{eqn:errr}) with respect to $\phi$ in order to determine the precision of a total spin readout in a neighborhood of $\phi=0$.

Finite-range one-axis twisting for quantum metrology applications was considered in Ref.\cite{PhysRevLett.112.103601} and implementation of layers of finite-range one-axis twisting in a parametrized spin-squeezing circuit was proposed in Ref.\cite{PhysRevLett.123.260505}. Ref.\cite{PhysRevA.94.010102} calculated the method of moments error of a Loschmidt echo readout for a two-dimensional finite-range, one axis twist-untwist protocol with respect to both interaction time and interaction range. The asymptotics of QFI and method of moments $J_{y}$ readout were calculated for finite-range, one-axis twist-untwist protocols in Ref.\cite{PhysRevResearch.4.013236}. The final untwist layer was allowed to have, in general, a different interaction time than the initial twist layer, but it was proven that when both interaction times scale as $t=O(K^{-1/2})$ (an interaction time domain in which both the QFI and the reciprocal method of moments error with $J_{y}$ rotation and $J_{y}$ readout scale as $O(NK)$), the asymptotically optimal  protocol is obtained with equal magnitude interaction times in the finite-range one-axis twisting layers, but with opposite interaction sign (i.e., a twist-untwist protocol).

\section{Background and examples \label{sec:back}}

In the limit of a large number of shots, the reciprocal of the error of the method of moments estimator $\tilde{\phi}$ of the rotation angle $\phi$ based on a total spin readout $\vec{m}\cdot \vec{J}$ has the form

\begin{equation}
   (\Delta \tilde{\phi})^{-2}_{\vec{n},\vec{m}}= { \left( \del_{\phi} \langle \vec{m}\cdot \vec{J} \rangle_{\ket{\psi_{\phi}(\vec{n})}} \right)^{2}\over \text{Var}_{\ket{\psi_{\phi}(\vec{n})}}\vec{m}\cdot \vec{J} }
   \label{eqn:errr}
\end{equation}
where $\vec{m} \in \mathbf{R}^{3}$ is a unit vector. For any $\phi$ and $\vec{m}$, the QFI is an upper bound for (\ref{eqn:errr}) \cite{ps}. Therefore, one can define an optimal method of moments protocol by maximizing (\ref{eqn:errr}) over $\vec{m}$ and $\phi$.

When the protocol has the form $\ket{\psi_{\phi}} = e^{-i\phi H}\ket{\psi_{0}}$ for some Hamiltonian $H$, it is often possible to simply take the $\phi \rightarrow 0$ limit of (\ref{eqn:errr}) by deriving expressions for the $\phi \rightarrow 0$ limits of the numerator and denominator and taking the quotient of these \cite{PhysRevLett.122.090503}. Unfortunately, this is not a valid approach for more general protocols such as $\ket{\psi_{\phi}}$ due to the possibility of zeros appearing in the numerator and denominator. A full analysis of (\ref{eqn:errr}) in the $\phi\rightarrow 0$ limit involves the off-diagonal spin $P$-function (coherent state symbol) for quartic polynomials in $J_{+}$ and $J_{-}$ \cite{kli,drum}. The amplitudes appearing in that expression can be evaluated by the generating function method \cite{PhysRevA.6.2211}. 

We now consider examples showing that at the extreme interaction times $t=0,\pi/2$, combining a twist-untwist protocol (\ref{eqn:ps}) with a total spin readout gives a reciprocal of the method of moments error (\ref{eqn:errr}) that achieves the maximal QFI over all SU(2) rotation orbits of the one-axis twisted state $e^{-itJ_{z}^{2}}\ket{\zeta=1}$ (at the same interaction times). First consider the classical sensing example of Yurke, et al. \cite{PhysRevA.33.4033}. The state $\ket{\sigma_{\phi}}=e^{-i\phi J_{z}}\vert \zeta =1\rangle$ is the $t=0$ limit both for a one-axis twisting orbit of $\ket{\zeta=1}$ and a twist-untwist protocol (\ref{eqn:ps}) (the sensing direction $J_{z}$ was chosen without loss of generality from the subspace spanned by $J_{z}$ and $J_{y}$). One finds that, e.g.,
\begin{equation}
     {\left( \del_{\phi}\langle J_{x} \rangle_{\ket{\sigma_{\phi}}}\right)^{2} \over \text{Var}_{\ket{\sigma_{\phi}}}J_{x}}= {{N^{2}\over 4}\sin^{2}\phi\over {N\over 4}\sin^{2}\phi}= N,
\end{equation}
i.e., the error of the method of moments estimate obtained from $J_{x}$ measurement achieves the SQL. This holds for all $\phi$, even $\phi$ for which the numerator and denominator go to zero (although such points are experimentally unstable). 

For the opposite extreme of $t=\pi/2$, consider the twist-untwist protocol state
\begin{eqnarray}
\ket{\psi_{\phi}}&=& e^{i{\pi\over 2}J_{z}^{2}}e^{-i\phi J_{x}}e^{-i{\pi\over 2}J_{z}^{2}}\ket{\zeta=1} \nonumber \\
&=& \cos {N\phi\over 2}\ket{\zeta=1}- \sin {N\phi\over 2}\ket{\zeta=-1}
\label{eqn:ghzexamp}
\end{eqnarray}
for even $N$ (if $N$ is odd, the rotation layer in (\ref{eqn:ghzexamp}) should be taken as $e^{-i\phi J_{y}}$, which gives the same form of final state as for $N$ even). Then $\text{Var}_{\ket{\psi_{\phi}}}J_{x}={N^{2}\over 4}\sin^{2}N\phi$ and $(\del_{\phi}\langle J_{x}\rangle_{\ket{\psi_{\phi}}} )^{2} = {N^{4}\sin^{2}N\phi \over 4}$ for all $\phi$. It follows that $(\Delta \tilde{\phi})^{2} =N^{-2}$ for all $\phi$ \cite{agarwal}. On the other hand, it is known that the state $e^{-i{\pi\over 2}J_{z}^{2}}\ket{\zeta=1}$ has maximal QFI $N^{2}$ for the rotation $e^{-i\phi J_{x}}$ ($e^{-i\phi J_{y}}$) for $N$ even (odd) \cite{PhysRevLett.57.13,PhysRevA.40.2417}. The fact that the twist-untwist protocol (\ref{eqn:ghzexamp}) allows to saturate Heisenberg limit using a total spin readout is remarkable because the state of the protocol is not even entangled for $\phi = {k\pi\over N}$, $z\in \mathbf{Z}$. However, the signal $(\del_{\phi}\langle J_{x}\rangle_{\ket{\psi_{\phi}}})^{2}$ vanishes at these points, maintaining a constant value of $(\Delta \tilde{\phi})^{2}$. The main point is that on the interval $(0,{\pi\over N})$, the state $\ket{\psi_{\phi}}$ follows a geodesic of the (collective) Bloch sphere, becoming distinguishable as fast as possible \cite{Volkoff2018distinguishability}. The present example also has the notable feature that the total spin readout $J_{x}$ is optimal regardless of the parity of $N$ due to the presence of the final untwist layer. By comparison, because the GHZ state $e^{-i{\pi\over 2}J_{z}^{2}}\ket{+}^{\otimes N}$ produced by one-axis twisting is aligned on the $x$-axis ($y$-axis) for $N$ even (odd), an optimal global parity readout as in (\ref{eqn:kkl}) must be aligned with the respective rotation depending on the parity of $N$ (viz., $J_{x}$ rotation and $X^{\otimes N}$ readout ($J_{y}$ rotation and $Y^{\otimes N}$ readout) is optimal for $N$ even (odd)).

The examples in this section raise the question of whether a twist-untwist protocol and total spin readout can be constructed that saturate the maximal one-axis twisting QFI for interaction times $t$ for $0<t< \pi /2$. In the next section, we determine the maximal one-axis twisting QFI for every interaction times and show that in the limit $N\rightarrow \infty$. the QFI takes the value $N(N+1)/2$ except on a set of interaction times of asymptotic measure zero. Then in Section \ref{sec:vvv}, we show that a twist-untwist protocol with $J_{x}$ rotation and $J_{x}$ readout gives a method of moments error with reciprocal that saturates this QFI in the $N\rightarrow \infty$ limit.

Finally, we note that our restriction to the one-axis twisting non-linear interaction is motivated by experiments in ultracold atomic gases in which one-axis twisted states are readily generated by manipulation of the two-body interaction. For some non-Gaussian states such as those in the twin Fock manifold spanned by $\ket{N/2,N/2}$, $\ket{N/2 -1,N/2 +1}$, $\ket{N/2 +1,N/2 -1}$, or ground states of the Bose-Hubbard dimer \cite{PhysRevLett.86.4431}, it is known to be possible, without implementing a twist-untwist protocol, to locally achieve Heisenberg scaling $O(N^{2})$ (near $\theta=0$) for the reciprocal of the method of moments error of a total spin readout or the square of a total spin readout \cite{PhysRevA.33.4033,PhysRevA.57.4004}. Such states are not accessible from the one-axis twisting dynamics of $\ket{\zeta=1}$ and we do not discuss these further.

\section{QFI for general interaction times\label{sec:gt}}

Before we identify twist-untwist protocols and total spin readouts that saturate the maximal QFI of the one-axis twisted state $e^{-itJ_{z}^{2}}\ket{\zeta=1}$ (over all rotation directions) for arbitrary interaction times $t$, it is important to understand how this maximal QFI scales with $N$ for each $t$. Through the QCRI, the maximal QFI value sets a lower bound for the error one can hope to achieve using the state $e^{-itJ_{z}^{2}}\ket{\zeta=1}$ for phase sensing. In fact, this maximal QFI is equal to the maximal QFI of (\ref{eqn:ps}) over all $\vec{n}$ because the QFI does not depend on the untwisting layer. Therefore the QFI is the same for both (\ref{eqn:ps}) and $e^{-i\phi \vec{n}\cdot \vec{J}}e^{-itJ_{z}^{2}}\ket{\zeta=1}$
with $\vec{n}=(\sin \xi \cos \theta , \sin \xi\sin\theta,\cos\xi)$. Specifically, the QFI is 
\begin{eqnarray}
&{}& 4\text{Var}_{e^{-itJ_{z}^{2}}\ket{\zeta=1}}(\sin \xi \cos \theta J_{x}+ \sin \xi \sin\theta J_{y}+\cos\xi J_{z})
\nonumber \\ &{}& =\sin^{2}\xi \left[ {N^{2}+N\over 2}+{N(N-1)\over 2}\cos 2\theta \cos^{N-2}2t  -N^{2}\cos^{2}\theta \cos^{2(N-1)}t \vphantom{N^{2}+N\over 2}\right]\nonumber \\
&{}& +N\cos^{2}\xi + \sin2\xi \left[ N(N-1)\sin \theta\cos^{N-2}t\sin t \right].  
\label{eqn:vvv}
\end{eqnarray}
 An analysis of the QFI for the special case $\vec{n}=(0,1,0)$ was given in \cite{ps}. The general features of the result of maximization over $\xi$ and $\theta$ were discussed in \cite{ps,PhysRevA.92.043622}. Note that the interaction time $t$, which can be alternatively written as $\chi T$ where $T$ is the real time and $\chi$ is the interaction strength, depends on $N$ through the $\chi$ factor. The various scalings of $t$ with $N$ allow asymptotic analyses of the maximal QFI. We proceed to use (\ref{eqn:vvv}) to derive the metrological phase diagram for rotation sensing with the states $e^{-itJ_{z}^{2}}\ket{\zeta=1}$. Figure \ref{fig:schem} shows the five regions of different maximal QFI scaling derived from the analysis of this section. 

\textbf{SQL scaling.} Consider the interaction time  $t=O\left( N^{-\alpha} \right)$ with $\alpha >1$, Using $\cos^{N}t \sim 1$ and $\sin t \sim {1\over N^{\alpha}}$, one finds that the bracketed term in (\ref{eqn:vvv}) scales as $O(N)$, the second term scales as $O(N)$, and the third term scales as $O(N^{2-\alpha})$.  Therefore, the first and second terms dominate, but $O(N)$ is the maximum QFI that can be obtained. Specifically, the QFI is asymptotically $N(\sin^{2}\xi\sin^{2}\theta +\cos^{2}\xi)$, from which one concludes that $\theta=\pm \pi/2$ and any $\xi$ gives the asymptotic global optimum. This result also holds for ultrashort interaction times $o(N^{-\alpha})$, $\alpha >1$. Logarithmic corrections to $t=O\left( N^{-\alpha} \right)$ do not change this asymptotic value.

\textbf{Sub-Heisenberg scaling.} Now consider $t=O( N^{-\alpha})$ for $1/2<\alpha<1$. Using $\cos^{N}t =1-{Nt^{2}\over 2}+O(t^{4})$ and related second-order polynomial approximations to the trigonometric monomials in (\ref{eqn:vvv}), one finds that the bracketed term in (\ref{eqn:vvv}) scales as $O(N^{3-2\alpha})$, the second term scales as $O(N)$, and the third term scales as $O(N^{2-\alpha})$. Therefore, the term in brackets dominates for large $N$ \cite{PhysRevLett.127.160501}, and one finds the asymptotically optimal $\xi=\pi/2$, $\theta=\pi/2$. In particular, one obtains a maximal QFI of $O(N^{5/3})$ at $t={24^{1/6}2^{2/3}\over 2N^{2/3}}$, the time at which one-axis twisting generates maximal spin squeezing \cite{PhysRevA.47.5138}. The dominance of the bracketed term persists for multiplicative logarithmic corrections to the $O(N^{-\alpha})$ scaling of $t$, and the QFI valued is correspondingly increased by a $\text{poly}(\log N)$ factor.

\textbf{Heisenberg scaling.}
There are two interaction time domains that allow to achieve $O(N^{2})$ QFI scaling which can be distinguished by their sensitivity to logarithmic corrections to the scaling of the interaction time. Consider first the case of an interaction time $t=O(N^{-\alpha})$, $0<\alpha <1/2$. It is clear that $\cos^{N}t\sim 0$, leading to an asymptotic QFI $N(N+1)\over 2$ for $\xi=\pi/2$ and arbitrary $\theta$. These interaction times correspond to oversqueezing  \cite{RevModPhys.90.035005,PhysRevA.47.5138, crphys}. For large $N$, the atomic Husimi $Q$ function for an oversqueezed one-axis twisting state has support on coherent states in a thin band surrounding the equator in the $xy$-plane, so it is not surprising that the maximal QFI is obtained independent of $\theta$. Further, the independence of asymptotically optimal QFI on the azimuthal angle implies that phase diffusion \cite{PhysRevLett.78.4675,PhysRevA.82.053603} during the call to the sensing parameter $\theta$ does not affect optimal QFI in this phase.

Multiplicative logarithmic corrections to this interaction time scaling do not affect the asymptotics. Further, this range of interaction times, which interpolates between the oversqueezed and GHZ regimes, best reflects the expected metrological performance of one-axis twisting for interferometry. In particular, 
\begin{equation}
    {2\over \pi}\int_{0}^{{\pi\over 2}}dt \, (\ref{eqn:vvv}) \sim \sin^{2}\xi {N^{2}+N\over 2} + o(N^{2})
    \label{eqn:avg}
\end{equation}
which, due to convexity of the QFI, is an upper bound on the QFI of the uniform statistical mixture (over interaction times) of the states (\ref{eqn:ps}). 

However, at smaller interaction times there is another scaling that leads to Heisenberg scaling of the QFI and encompasses the transition from sub-Heisenberg scaling to Heisenberg scaling. Consider the interaction time $t=O\left( {(\log N)^{\beta}\over \sqrt{N}}\right)$ with $\beta \in \mathbf{R}$. For $\beta =0$, taking $t={c\over \sqrt{N}}$ results in an asymptotic QFI $N^{2}{1-e^{-2c^{2}}\over 2}$ at the optimal angles $\xi = \theta = \pi/2$. The interaction time regime $t={c\over \sqrt{N}}$ is also where the properties of the twist-untwist protocol state (\ref{eqn:ps}) with $\vec{n}=(0,1,0)$ have been most fully explored \cite{PhysRevResearch.4.013236,PhysRevLett.116.053601}. Although those works analyzed the optimal sensing direction $\vec{n}=(0,1,0)$, they only showed that a reciprocal method of moments error for the total spin $J_{y}$ comes close to saturating the QFI (about 85\%, asymptotically in $N$). In the present work, we show in Section \ref{sec:vvv} that for arbitrarily small $\phi$, there is a total spin readout with a reciprocal method of moments error that actually saturates the maximal QFI. 

For $\beta >0$ in the interaction time, the Heisenberg scaling of the asymptotic QFI is converted to the asymptotic Heisenberg scaling $N(N+1)/2$ discussed above. However, for $\beta <0$, the optimal Heisenberg scaling is converted to optimal QFI of the form $O(N^{2}(\log N)^{2\beta})$. The proof of this fact closely follows the discussion of sub-Heisenberg scaling above, simply noting that an interaction time $t=O\left( {(\log N)^{\beta} \over \sqrt{N}}\right)$, $\beta <0$, still allows one to utilize the MacLaurin series for $\cos^{N}t$ at lowest nontrivial order. It follows that an interaction time scaling $t=O\left( {(\log N)^{\beta}\over \sqrt{N}}\right)$ encompasses the transition to Heisenberg scaling.

\textbf{Heisenberg limit.}
From (\ref{eqn:vvv}), one sees that the global maximum $N^{2}$ of the QFI is obtained for $t={\pi\over 2}$ and for a rotation about $x$-axis ($N$ even) or $y$-axis ($N$ odd). Combining this fact with the result from the Heisenberg scaling region, one concludes that the asymptotically optimal sensing direction for interaction times $c/N^{\alpha}$ with $\alpha \le 1/2$ is in the linear span of $J_{x}$ and $J_{y}$. The width of the GHZ region around $t={\pi\over 2}$ is $O(1/\sqrt{N})$. This can be proved by showing that this neighborhood of ${\pi\over 2}$ interpolates between the Heisenberg limit $N^{2}$ and the ${N^{2}/2}$ QFI asymptotic calculated above for $t=O(N^{-\alpha})$, $0<\alpha <1/2$. Specifically, consider $t={\pi\over 2}-{c\over \sqrt{N}}$ for constant $c>0$.  From the fact that $\cos^{N}\left( {\pi \over 2}-{c\over \sqrt{N}} \right) \rightarrow 0$, one finds that the third term (\ref{eqn:vvv}) and the third term in brackets in the first line of (\ref{eqn:vvv}) do not contribute, while the second term of (\ref{eqn:vvv}) is $O(N)$. From the fact that $\cos^{N-2}\left( \pi-{2c\over \sqrt{N}} \right) \sim \pm e^{-2c^{2}}$ (top (bottom) sign holds for $N$ even (odd)), one finds that (\ref{eqn:vvv}) is asymptotically
\begin{equation}
    \sin^{2}\xi \left[ {N^{2}\over 2}\left( 1\pm \cos2\theta e^{-2c^{2}} \right) + {N\over 2}\left(1\mp \cos2\theta e^{-2c^{2}} \right) \right].
\end{equation}
Taking $\xi = \pi/2$, $\theta = 0$ for $N$ even and $\xi = \pi/2$, $\theta=\pi/2$ for $N$ odd, this is the sought  interpolation between $N^{2}$ (for small $c$) and ${N^{2}+N\over 2}$ (for large $c$).  It is also worth noting that for constant interaction times $t={\pi \over m}$ ($m>2$) defining discrete superpositions of equidistant coherent states \cite{PhysRevA.47.5024}, the asymptotically maximal QFI $N(N+2)/2$ is obtained with $\xi = \pi/2$ and arbitrary $\theta$. This can be seen from the exponentially small contributions from $\cos t$ and $\cos 2t$ in (\ref{eqn:vvv}) for constant $0<t<\pi/2$.

The above analyses of the QFI scaling as a function of interaction time allows to draw the conclusion that the protocol (\ref{eqn:ps}) (or any protocol obtained from it by applying a $\phi$-independent unitary) achieves its optimal metrological usefulness on the circle $\xi=\pi/2$, $\theta \in [-\pi,\pi)$, asymptotically in $N$. This useful corollary is not obvious from examination of a phase space distribution for (\ref{eqn:ps}), which is not aligned to a fixed direction on the sphere over all interaction times.  Finally, we note that (\ref{eqn:avg}) suggests the hypothesis that the angles $\xi = \pi/2$, $\theta=0$ are asymptotically almost everywhere optimal for all interaction times $t\in [0,\pi/2]$, and this QFI value is ${N^{2}+N\over 2}$. This hypothesis is true, and we provide a formal statement and proof.
\begin{lemma}
\label{lem:11}
For any $0<\epsilon,\epsilon'<\pi/2$, there exists an $N$ such that: 1. the maximum of (\ref{eqn:errr}) on an $N$-dependent interaction time interval $T_{N}\subset [0,\pi/2]$ is within $\epsilon'$ of ${N(N+1)/2}$, and 2. $\vert [0,\pi/2] \setminus T_{N} \vert <\epsilon$. 
\end{lemma}

\textit{Proof}. Consider an interaction time of the form ${\pi\over 2}-{c\over N^{\beta}}$ with $0<\beta<1/2$. In this case, $\cos^{N}\left( {\pi\over 2}-{c\over N^{\beta}}\right) \rightarrow 0$ and $\cos^{N}\left( \pi-{2c\over N^{\beta}}\right) \rightarrow 0$; therefore, the first term in brackets in (\ref{eqn:vvv}) dominates for large $N$. Depending on $\epsilon '$, we then take $N$ large enough that $\xi = {\pi/2}$ and arbitrary $\theta$ gives a value within $\epsilon'$ of $N(N+1)/2$. On the other hand, we derived above that for $\xi=\pi/2$, $\theta$ arbitrary, and interaction time $t=O(N^{-\alpha})$, $0<\alpha <1/2$, one can choose $N$ sufficiently large that (\ref{eqn:vvv}) is within $\epsilon'$ of $N(N+1)/2$. It follows that for $\epsilon'>0$ and any $0<\alpha,\beta<1/2$ and any constants $c,c'$, one can choose a large enough $N$ such that (\ref{eqn:vvv}) is within $\epsilon'$ of $N(N+1)/2$ on the interval $T_{N}=[{c\over N^{\alpha}},{\pi\over 2}-{c'\over N^{\beta}}]$ for $\xi=\pi/2$ and arbitrary $\theta$. Depending on $\epsilon$, one then further increases $N$ such that the difference between $[0,\pi/2]$ and this interval has measure $\epsilon$. $\square$

The log scale for the interaction time in Figure \ref{fig:schem} is used because it better illustrates the different phases. On a linear scale, the QFI is clearly seen to take the value $N(N-1)/2$ for almost the whole interaction time domain, as expected from Lemma \ref{lem:11}.

\section{Twist-untwist protocols\label{sec:vvv}}

We now show that for $\phi\rightarrow 0$, the twist-untwist protocol with $J_{x}$ rotation and a $J_{x}$ measurement gives an error for the method of moments estimator of $\phi$ that is asymptotically equal to $N^{2}/2$ for interaction times $t=O(N^{-\alpha})$, $0<\alpha<1/2$. This conclusion holds regardless of the parity of $N$. Due to Lemma \ref{lem:11}, this result implies that a twist-untwist protocol with total spin readout saturates the QCRI for one-axis twisting asymptotically for almost all interaction times (because the metrological phase diagram  Fig.\ref{fig:schem} degenerates to the pink Heisenberg scaling phase).

\begin{figure*}[t!]
\centering
\includegraphics[scale=0.55]{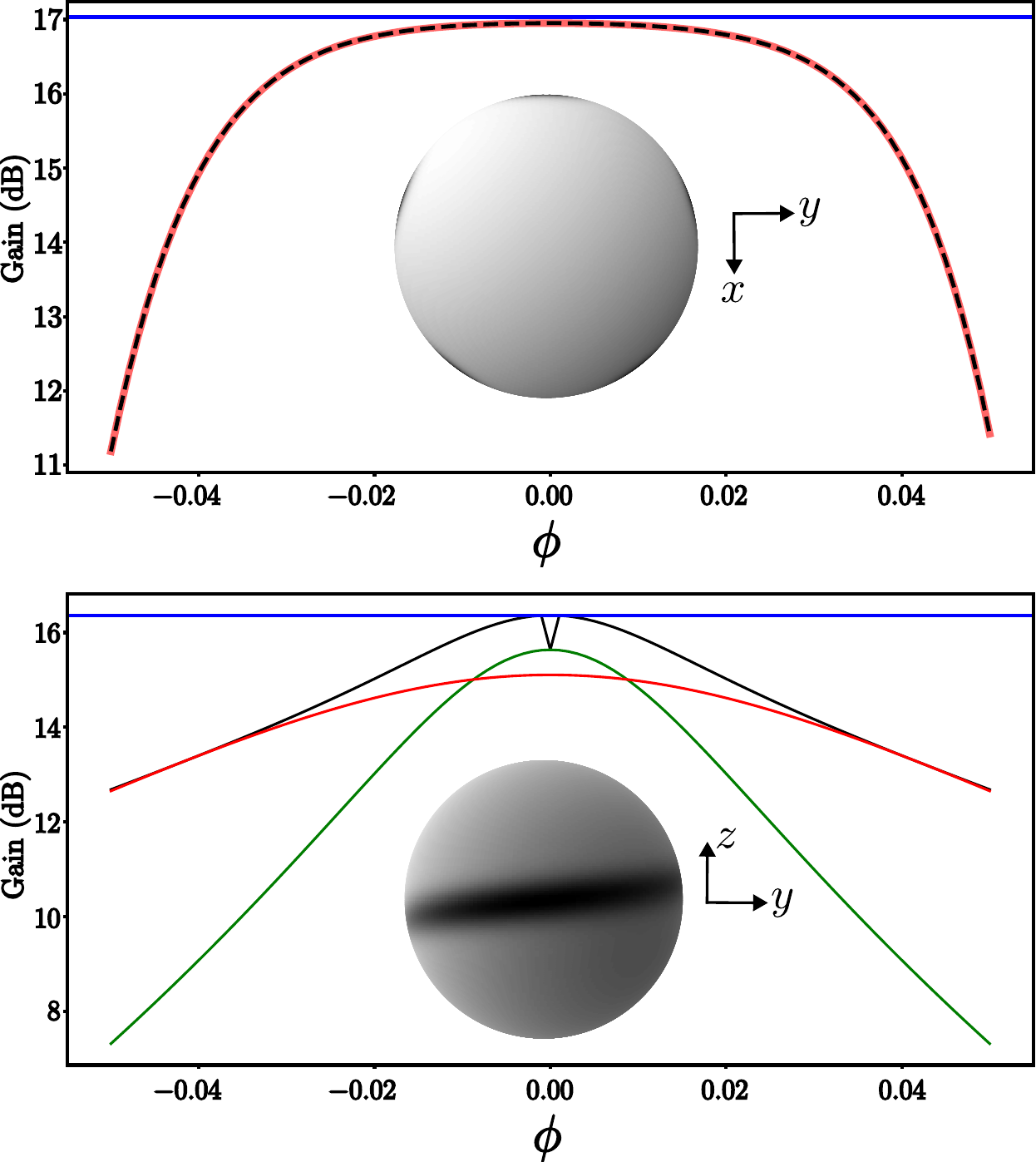}
    \caption{(Top) (Blue) Maximal QFI over $\vec{n}$ at interaction time $N^{-1/10}$ (recall from Section \ref{sec:gt} that for large $N$, the optimal $\vec{n}$ for this time domain can be anywhere in $xy$-plane). (Dashed black) Maximal value of (\ref{eqn:errr}) over $\vec{m}$ with fixed $J_{x}$ rotation. (Red) Reciprocal method of moments error with fixed $J_{x}$ rotation and $J_{x}$ readout. $J_{y}$ readout is less than the SQL and is not shown. (Bottom) (Blue) Maximal QFI over $\vec{n}$ at interaction time $N^{-1/2}$ (recall from Section \ref{sec:gt} that for large $N$, the optimal $\vec{n}$ is $(0,1,0)$  for this time domain). (Black) Maximal value of (\ref{eqn:errr}) with fixed $J_{y}$ rotation. (Red) Reciprocal method of moments error with fixed $J_{y}$ rotation and $J_{x}$ readout. (Green) Reciprocal method of moments error with fixed $J_{y}$ rotation and $J_{y}$ readout. $N=100$ throughout. Insets are spin Husimi $Q$-function for the probe state after the first one-axis twisting is applied.
}
    \label{fig:ppp}
\end{figure*}

To proceed, note that the denominator of (\ref{eqn:errr}) involves the derivative of the total spin signal, which can be written 
\begin{eqnarray}
    &{}&\del_{\phi}\langle\vec{m}\cdot\vec{J}\rangle_{\vert \psi_{\phi}(\vec{n})\rangle}=-2\text{Im}\langle \zeta =1\vert e^{itJ_{z}^{2}}\vec{n}\cdot \vec{J}e^{i\phi \vec{n}\cdot \vec{J}}e^{-itJ_{z}^{2}}\vec{m}\cdot\vec{J}\vert \psi_{\phi}\rangle
    \label{eqn:siggen}
\end{eqnarray}
To analyse the $\phi\rightarrow 0$ behavior of this expression, one should calculate the MacLaurin expansion at least to $O(\phi)$ because the value of the expression may vanish at $\phi=0$. We now show that for $t=O(N^{-\alpha})$, $0<\alpha<1/2$, $\max_{\vec{n},\vec{m}} (\Delta \tilde{\phi})^{2}\vert_{\phi=0} \sim {N(N+1)\over 2}$ is obtained by taking $\vec{n}=\vec{m}=(1,0,0)$, i.e., $J_{x}$ rotation and $J_{x}$ readout. First note that for these values of $\vec{n},\vec{m}$, (\ref{eqn:siggen}) is zero at $\phi=0$, so the $O(\phi^{2})$ contribution to the denominator of (\ref{eqn:errr}) is the square of the $O(\phi)$ contribution to (\ref{eqn:siggen}). One finds that to $O(\phi)$
\begin{eqnarray}
    &{}& \del_{\phi}\langle J_{x} \rangle_{\vert \psi_{\phi,\vec{n}}\rangle}= -N\phi \langle \zeta=1 \vert e^{itJ_{z}^{2}}J_{x}^{2}e^{-itJ_{z}^{2}}\vert \zeta=1\rangle\nonumber \\
    &{}& +2\phi \langle\zeta=1\vert e^{itJ_{z}^{2}}J_{x}e^{-itJ_{z}^{2}}J_{x}e^{itJ_{z}^{2}}J_{x}e^{-itJ_{z}^{2}}\vert \zeta=1\rangle.
    \label{eqn:rfrf}
\end{eqnarray}
The first line can be evaluated by a standard computation and the second line can be evaluated using Kitagawa-Ueda formulas and a table of third moments of the $J_{\pm}$ operators in spin coherent states (see \ref{sec:app1}). The result is in (\ref{eqn:resres}). Taking $t=O(N^{\alpha})$, $0<\alpha<1/2$ gives 
\begin{equation}
    \left( \del_{\phi}\langle J_{x}\rangle_{\ket{\psi_{\phi}((1,0,0))}} \right)^{2} \sim { N^{6-4\alpha}\phi^{2}\over 16}.
    \label{eqn:tqtq}
\end{equation}
 Unfortunately, the $O(\phi^{2})$ contribution to the variance cannot be obtained from the same set of operator moments as (\ref{eqn:rfrf}), so a further calculation gives
 \begin{eqnarray}
     \text{Var}_{\ket{\psi_{\phi}((1,0,0))}}J_{x}&=&{N\phi^{2}\over 4}\langle e^{itJ_{z}^{2}}J_{x}^{2}e^{itJ_{z}^{2}}\rangle_{\ket{\zeta=1}} + \phi^{2}\langle e^{itJ_{z}^{2}}J_{x}e^{-itJ_{z}^{2}}J_{x}^{2}e^{itJ_{z}^{2}}J_{x}e^{-itJ_{z}^{2}}\rangle_{\ket{\zeta=1}}\nonumber \\
     &{}& -N\phi^{2}\langle e^{itJ_{z}^{2}}J_{x}e^{-itJ_{z}^{2}}J_{x}e^{itJ_{z}^{2}}J_{x}e^{-itJ_{z}^{2}}\rangle_{\ket{\zeta=1}}+O(\phi^{4}).
     \label{eqn:ghgh}
 \end{eqnarray}
 The first and last terms of (\ref{eqn:ghgh}) can be evaluated using  expressions from (\ref{eqn:rfrf}), whereas the second term requires tabulation of fourth moments of $J_{\pm}$ in spin coherent states (\ref{sec:app2}).  The final asymptotic result for the variance is
 \begin{equation}
     \text{Var}_{\ket{\psi_{\phi}((1,0,0))}}J_{x}\sim  {N^{4-4\alpha}\phi^{2}\over 8}.
     \label{eqn:varzro}
 \end{equation}
 From the quotient of (\ref{eqn:tqtq}) and (\ref{eqn:varzro}), one obtains the result
 \begin{equation}
    {\left( \del_{\phi}\langle J_{x}\rangle_{\ket{\psi_{\phi}((1,0,0))}} \right)^{2}\over \text{Var}_{\ket{\psi_{\phi}((1,0,0))}}J_{x} } \big\vert_{\phi=0} \sim {N^{2}\over 2}.
    \label{eqn:mmjxrd}
 \end{equation}
 This equation proves that except for a set of interaction times of asymptotic measure zero, the $\ket{\psi_{\phi}((1,0,0))}$ twist-untwist protocol with $J_{x}$ total spin readout saturates the quantum Cram\'{e}r-Rao bound for one-axis twisting in the limit of large $N$ (see Fig. \ref{fig:ppp} (top) for graphical depiction for finite $N$). One can also numerically verify for interaction times not in the pink phase of Fig. \ref{fig:schem} that for large $N$, there is always a $\phi$ such that the optimal method of moments error for a twist-untwist protocol (\ref{eqn:errr}) saturates the one-axis twisting QCRI. An example is shown in Fig. \ref{fig:ppp} (bottom) for $t=N^{-1/2}$, where an arbitrarily small nonzero $\phi$ allows the maximal (\ref{eqn:errr}) to saturate the QFI (there is a discontinuity at $\phi=0$).  Such optimal twist-untwist protocols for small interaction times are relevant for utilizing total spin readouts for sensing $\phi$ in finite-size atomic ensembles which may not be able to reach interaction times for oversqueezing.
 
 Because the one-axis twisting QFI is constant on a unitary path (defined only by the interaction time $t$ and the rotation generator parameters $\xi$ and $\theta$ defining $\vec{n}$), the one-axis twisting metrological phase diagram (\ref{fig:schem}) and Lemma \ref{lem:11} are globally valid for all sensing angles $\phi$. However, the results in this section only indicate that the reciprocal of the method of moments error (\ref{eqn:errr}) saturates the maximal one-axis twisting QFI when the former is evaluated at $\phi=0$. To extend the twist-untwist protocol (\ref{eqn:ps}) to a protocol that allows one to asymptotically saturate the one-axis twisting QCRI using the method of moments estimation from total spin readout for any rotation angle $\phi$, it suffices to add a single controllable rotation after the call to the parameter $\phi$:
 \begin{equation}\ket{\tilde{\psi}_{\phi}(\vec{n})}=e^{itJ_{z}^{2}}Re^{-i\phi \vec{n}\cdot \vec{J}}e^{-itJ_{z}^{2}}\ket{\zeta =1}\label{eqn:ps2}\end{equation}
 where $R$ is a parametrized $SU(2)$ operation which can be optimized over while measuring the total spin in the optimal direction. This optimization is analogous to the realigning pulses in a trapped Bragg interferometry sequence \cite{PhysRevA.103.L061301}. Since $R$ is independent of $\phi$, $\ket{\tilde{\psi}_{\phi}(\vec{n})}$ has the same QFI as (\ref{eqn:ps}) or the one-axis twisted state $e^{-i\phi \vec{n}\cdot \vec{J}}e^{-itJ_{z}^{2}}\ket{\zeta=1}$. Mathematically, it is clear that taking $R=e^{i\phi'\vec{n}\cdot \vec{J}}$ is the same as evaluating (\ref{eqn:errr}) at $\phi'-\phi$. Scanning $\phi'$ then allows to obtain the $\phi=0$ method of moments error, but for any $\phi$.   In a Mach-Zehnder interferometry sequence, one takes $\vec{n}\cdot \vec{J} = J_{x}$
 or $J_{y}$, and the phase $\phi$ is imprinted as a phase shift of modes via $e^{-i\phi J_{y}(J_{x})}=e^{-i{\pi\over 2}J_{x}(-J_{y})}e^{i\phi J_{z}}e^{i{\pi\over 2}J_{x}(-J_{y})}$, so for a general phase $\phi$, a protocol that asymptotically saturates the one-axis twisting QCRI is
 \begin{equation}e^{itJ_{z}^{2}}e^{i\phi'J_{y}(J_{x})}e^{-i{\pi\over 2}J_{x}(-J_{y})}e^{-i\phi \vec{n}J_{z}}e^{i{\pi\over 2}J_{x}(-J_{y})}e^{-itJ_{z}^{2}}\ket{\zeta =1}.\label{eqn:ps3} 
 \end{equation}
 with $J_{x}$ or $J_{y}$ chosen according to the asymptotically optimal $\xi,\theta$ pair derived in Section \ref{sec:gt} for any given interaction time $t$.
 
 One undesirable aspect of the optimal twist-untwist protocol in the Heisenberg scaling phases and in the  Heisenberg limit phase near interaction time $\pi/2$, is the fact that the optimal twist-untwist protocol utilizes a signal $\langle J_{x}\rangle_{\ket{\psi_{\phi}}}$ which has zero derivative with respect to $\phi$ at $\phi=0$. In practice, this behavior leads to unstable behavior of the method of moments estimator of $\phi$ in a neighborhood of zero. A potential route to circumvent this aspect of the optimal twist-untwist protocol is to utilize (\ref{eqn:ps2}). For example, considering $t=O(N^{-\alpha})$, $0<\alpha <1/2$, the optimal $\phi$ rotation direction in the twist-untwist protocol is $\vec{n}=(1,0,0)$, and the optimal readout is $\vec{m}=(1,0,0)$, but according to (\ref{eqn:ps2}), one lets $R$ be a tunable non-identity rotation. Because the signal would have non-zero derivative, the experiment can tune $R$ to be as close to the identity as possible while still getting Heisenberg scaling for the method of moments estimate of $\phi$.
 
 \section{Finite-range one-axis twisting \label{sec:foat}}

We now consider spin squeezing operations on an ensemble of distinguishable two-level atoms, such as Rydberg atoms in an optical lattice. Consider $N+2$ spin-1/2 particles interacting on a 1-D lattice with periodic boundary conditions according to the interaction Hamiltonian 
\begin{equation}
    H_{K}={1\over 4}\sum_{j=1}^{N+2}\sum_{i=j-K\atop i\neq j}^{j+K}Z_{i}Z_{j}
\end{equation}
where $Z_{i}$ is the Pauli $Z$ matrix on site $i$. Such an interaction corresponds to a uniform coupling of  spins over a range $K$. We take $N\equiv 0\,\text{ mod }\,2$  and note that \begin{equation}H_{N/2}+{1\over 4}\sum_{j=1}^{N}Z_{j}Z_{j+1+{N\over 2}\text{mod} N}= 2J_{z}^{2}\end{equation} with $J_{z}$ in a $j={N+2\over 2}$ representation of $SU(2)$. Surprisingly, removing the antipodal interaction and considering just the long range one-axis twisting interaction $H_{N/2}$ produces a metrological phase diagram that qualitatively differs from Fig. \ref{fig:schem}. 

\begin{center}
\begin{table}[t]
\caption{Metrological phases for finite-range one-axis twisting. The second, third, and fourth columns show the scaling of the interaction time that defines the metrological phase.}
\centering
\begin{tabular}{p{0.12\linewidth}p{0.25\linewidth}p{0.35\linewidth}p{0.25\linewidth}}
\hline
 Interaction range& SQL& SQL-Heisenberg interpolation& Heisenberg \\
\hline
$1\le K \le {N\over 4}$ & $\Omega\left( {(\log K)^{\beta} \over K^{\alpha}} \right)$; $\alpha >1/2$, $\beta \ge 0$ & $\Omega\left( (\log K)^{\beta} \over K^{\alpha} \right)$; $0<\alpha \le 1/2$, $\beta > 0$  & $\Omega(K^{-1/2})$, $K=\lambda N$ \\
{}&{}&$O\left(1\right)$, $\pi/2$ included&{}\\
\hline
$K= {N\over 2}$& $\Omega\left( {(\log N)^{\beta} \over N^{\alpha}} \right)$; $\alpha >1/2$, $\beta \ge 0$& $\Omega\left( {(\log N)^{\beta} \over N^{\alpha}} \right)$; $0<\alpha \le 1/2$, $\beta > 0$&$\Omega\left( N^{-1/2}\right)$\\
{}&${\pi \over 2}-O(N^{-1/2})$ (doubled SQL)&$O\left( 1\right)$, $\pi/2$ not included\\
\hline
\end{tabular}
\label{tab:frphasediagram}
\end{table}
\end{center}

We start by deriving a metrological phase diagram for finite-range one-axis twisting generated by $H_{K}$ in the case $1\le K\le {N\over 4}$. The interaction times corresponding to the various QFI scalings are summarized in Table \ref{tab:frphasediagram}, and in this section we provide details on these asymptotic expressions. For this short finite-range interacting case, the QFI on the path $e^{-i\phi \vec{n}\cdot \vec{J}}$ is in (\ref{eqn:smallk}) in \ref{sec:qfifr}. From (\ref{eqn:smallk}) one can see that the QFI for finite range one-axis twisting is determined by rational trigonometric functions, with growth determined by the function $\cos^{K}mt$, $m=1,2$. We therefore consider the interaction time to scale in various ways with respect to $K$, and take the large $K$ limit, always noting that $N$ must be taken greater than or equal to $4K$. One should note that if the range $K$ is fixed, there is no hope for $O((N+2)^{2})$ scaling of the QFI. From (\ref{eqn:smallk}), one sees that the $O((N+2)^{2})$ term simplifies to zero, regardless of the readout $\vec{n}\cdot \vec{J}$. Therefore, any $O(N^{2})$ scaling would come from taking $K$ to scale linearly with $N$ (with slope $\le {1/4}$).

\textbf{SQL scaling.} There is a single metrological phase exhibiting strict SQL scaling when $1\le K\le {N\over 4}$. Consider $t(K)={c(\log K)^{\beta}\over K^{\alpha}}$ with $\alpha >1/2$, $\beta \ge 0$. We have $\cos^{K}t(K)\sim 1$, and applying L'Hopital's rule to (\ref{eqn:smallk}), one obtains that the QFI is 
\begin{equation}
    (N+2)(\cos^{2}\xi -\sin^{2}\xi\cos 2\theta)
\end{equation}for large $K$, which implies that taking $\xi=0$ (the path generated by $J_{z}$) or $\xi={\pi\over 2}$, $\theta=\pi$ (the path generated by $J_{y}$) achieves SQL. This is consistent with the two paths that give QFI equal to $N$ for the $t=0$ non-spin-squeezed state $\ket{+}^{\otimes N}$.

\begin{figure}[t!]
\begin{center}
\includegraphics[scale=0.6]{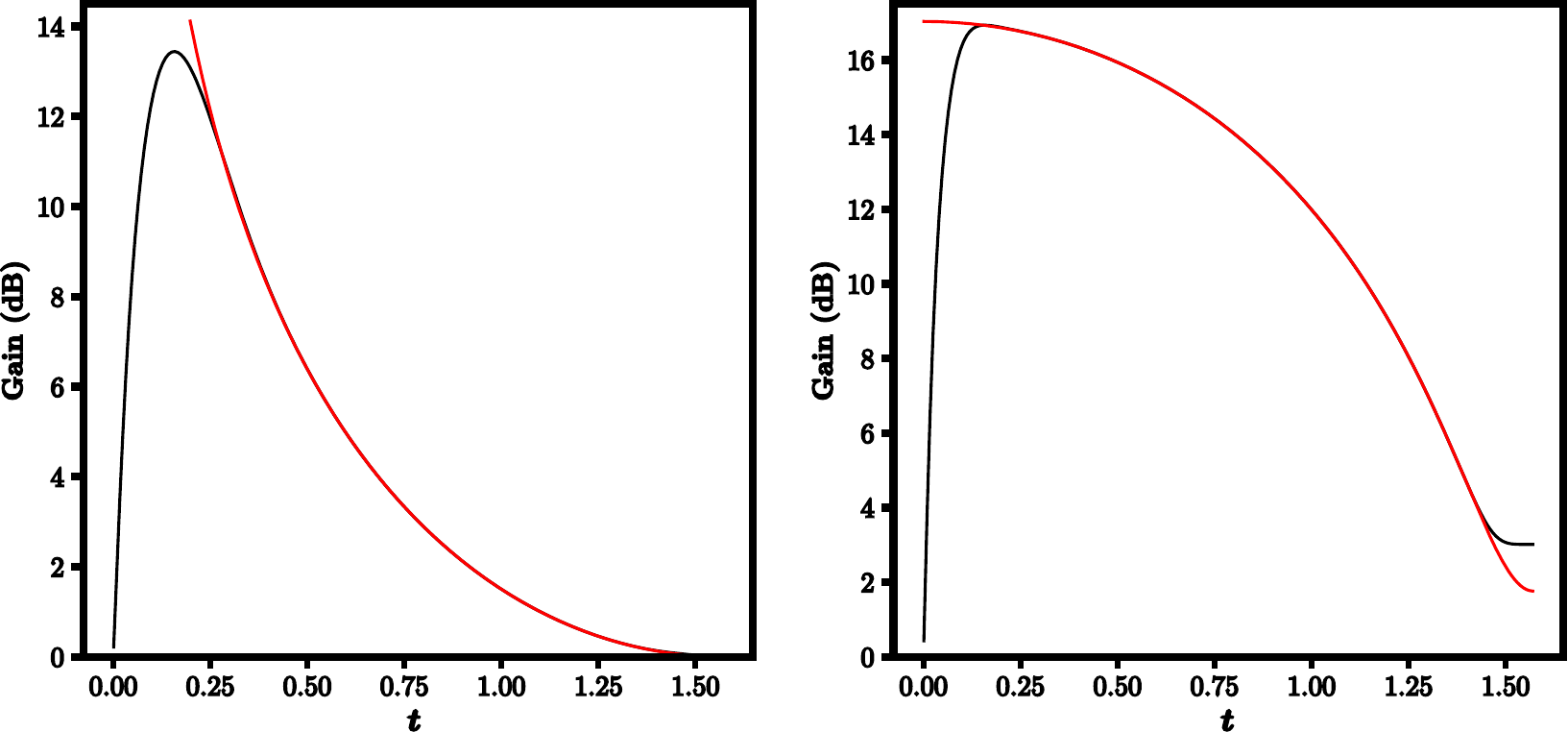}
 \caption{\label{fig:fig3}(Left) Red: (\ref{eqn:inter1}) for $t\in (0,\pi/2)$ for $N+2=100$, Black: ${4\over N+2}\text{max}_{\vec{n}}\text{Var}\vec{n}\cdot \vec{J}$ in the state $e^{-itH_{K}}\ket{+}^{\otimes N+2}$ (reported in decibels; we compute the maximum numerically from the analytical expression for the variance in \ref{sec:qfifr}) on $t\in (0,\pi/2)$, with $K=(N+2)/4$ and $N+2=100$ (Right) Red: (\ref{eqn:largescaleform}) for $t\in (0,\pi/2)$ for $N+2=100$, Black: ${4\over N+2}\text{max}_{\vec{n}}\text{Var}\vec{n}\cdot \vec{J}$ (in decibels) in the state $e^{-itH_{K}}\ket{+}^{\otimes N+2}$ on the same interaction time interval, with $K=(N+2)/2$ and  $N+2=100$.
}
\end{center}
\end{figure}

\textbf{Heisenberg-SQL interpolation.} We have already established that when $1\le K\le {N\over 4}$, $O(N^{2})$ scaling of the QFI cannot occur for fixed $K$. However, there is the possibility that the QFI scales as $O(NK^{2\alpha})$ for large $K$, which induces Heisenberg scaling for $\alpha = 1/2$ and $K=\lambda N$ (with $0<\lambda< {1\over 4}$ because we are considering $1\le K\le {N\over 4}$).  To investigate this possibility, consider any scaling $t(K)$ such that $\cos^{K}t(K) \rightarrow 0$, e.g., $t(K)={c(\log K)^{\beta}\over K^{\alpha}}$, $0<\alpha \le 1/2$, $\beta >0$, or constant scaling $t=c\in (0,{\pi\over 2}]$. In this case, the QFI is
\begin{equation}
    {N+2\over \sin^{2}t(K)}{\sin^{2}\xi} + (N+2){\cos^{2}\xi},
    \label{eqn:inter1}
\end{equation}
which for large $K$ is maximized by $\xi={\pi\over 2}$ regardless of $\theta$, with corresponding value ${N+2\over \sin^{2}t(K)}$. The scaling in (\ref{eqn:inter1}) is shown in Fig. \ref{fig:fig3}. For large $K$, this functional form fills out the entire metrological phase diagram. Using form of $t(K)$ above, one notes that the QFI on a rotation path defined by $\vec{n}$ in the $xy$-plane is asymptotically ${(N+2)K^{2\alpha}\over c^{2}(\log K)^{2\beta}}$ for large $K$, which is corresponds to Heisenberg scaling if an aspect ratio $0<\lambda\le {1\over 4}$ is defined where $K=\lambda N$. By contrast, taking a constant $t$ (which is interpolated by the form of $t(K)$) gives SQL scaling. Therefore, these interaction time scalings interpolate between Heisenberg scaling and SQL scaling of the QFI.  

\textbf{Heisenberg scaling.}
Because logarithmic corrections to $t=\Omega(K^{-1/2})$ scaling are already covered in the analyses above, it only remains to analyze $t={c\over \sqrt{K}}$ with $c>0$, i.e., $\beta=0$ and $\alpha =1/2$ in $t(K)$. To do this, we neglect the terms that do not scale with both $N$ and $K$ (the $O(N^{2})$ term vanishes identically, so this actually amounts to neglecting $O(NK^{0})$ terms). The QFI is
\begin{equation}
    {(N+2)K\sin^{2}\xi \over 4}\left( {1-e^{-2c^{2}}\over c^{2}} - 2e^{-2c^{2}}  + \cos 2\theta \left( {e^{-2c^{2}}-e^{-4c^{2}}\over c^{2}}-2e^{-2c^{2}} \right) \right)
    \label{eqn:bestshort}
\end{equation}
maximized for $\xi=\theta=\pi/2$ regardless of $c$. This $O((N+2)K)$ scaling of the QFI is the best possible for short, finite-range one axis twisting, and corresponds to Heisenberg scaling when an aspect ratio $\lambda$ is chosen. Calculations of the QFI (optimized over all directions) for two-dimensional, finite-range one-axis twisting with a $1/r_{ij}^{6}$ potential relevant to lattice Rydberg atoms further confirm that, unlike the case for one-axis twisting, a QFI greater than SQL does not persist beyond a short regime of interaction times  \cite{PhysRevA.94.010102}. The same work also shows an asymptotic scaling $NK^{\gamma}$ of the QFI, where $\gamma \approx 1.94$  and where $K$ represents the range of the interaction as a multiple of the lattice spacing. Along with (\ref{eqn:bestshort}), which shows an $O(NK)$ scaling in the one-dimensional setting, this suggests a dimensionality dependence of the maximal QFI $\propto NK^{O(d)}$ for short finite-range one-axis twisting in dimension $d$.

Finally, we derive a metrological phase diagram for finite-range one-axis twisting generated by $H_{K}$ in the case $ {N\over 4}<K\le {N\over 2}$. For this long finite-range interacting case, the QFI on the path $e^{-i\phi \vec{n}\cdot \vec{J}}$ is in (\ref{eqn:bigk}) in \ref{sec:qfifr}. Unlike the case for $ {N\over 4}<K\le {N\over 2}$, the exponents in the rational trigonometric functions are not just multiples of $K$, so it is unclear how to choose the scalings of the interaction times. A totally general analysis would define an aspect ratio $K=\lambda N$ with ${1\over 4}<\lambda \le {1\over 2}$ and for each $\lambda$ consider interaction time scaling with $N$. However, the most interesting question concerning quantum metrology with long finite-range one-axis twisting interaction is whether the metrological phase diagram interpolates to that of the $J_{z}^{2}$ interaction (i.e., the bosonic case) for $\lambda = 1/2$. Therefore, we take $\lambda = 1/2$ and analyze the interaction time scaling with $N$.

\textbf{SQL scaling.} There are two disjoint metrological phases that correspond to SQL scaling. First, consider interaction time scaling $t(N)={c(\log N)^{\beta}\over N^{\alpha}}$, $\alpha >{1/2}$, $\beta \ge 0$. We have $\cos^{N}t(N)\sim 1$ and applying L'Hopital's rule to (\ref{eqn:bigk}), one obtains that the QFI is
\begin{equation}
    (N+2)\left( {5\sin^{2}\xi\over 2} + {3\sin^{2}\xi \cos 2\theta \over 2}  +{\cos^{2}\xi} -4\sin^{2}\xi\cos^{2}\theta \vphantom{{5\sin^{2}\xi\over 2}}\right).
\end{equation}
This is maximized for $\xi=0$ ($J_{z}$ generator), or $\xi=\pi/2$, $\theta=\pi/2$ ($J_{y}$ generator), with maximal value $(N+2)$ in both cases. Second, consider $t={\pi\over 2}$ which gives QFI
\begin{equation}
    (N+2)\left( {3\sin^{2}\xi\over 2} + {\sin^{2}\xi \cos2\theta \over 2} + \cos^{2}\xi \right).
\end{equation}
This obtains maximum $2(N+2)$ for $\xi={\pi\over 2}$, $\theta=0$. This ``doubled SQL'' scaling persists for interaction time scaling $t(N)={\pi\over 2}-{c\over \sqrt{N}}$, $c>0$. Note that if $N$ is odd, the optimal $\theta$ is $\pi/2$ instead of $0$.

\textbf{Heisenberg-SQL interpolation.} Consider interaction time $t(N)$ scaling with $N$ in such a way that $\cos^{N}t(N)\rightarrow 0$. For example, $t(N)={c(\log N)^{\beta}\over N^{\alpha}}$ with $0<\alpha \le {1\over 2}$ and $\beta>0$, or constant $t=c\in (0,{\pi\over 2}]$. In this case, (\ref{eqn:bigk}) is asymptotically independent of $\theta$ and the QFI is then given by
\begin{equation}
{\sin^{2}\xi\over 8}\left( \cos^{2}t(N)(N^{2}-4)+2(N+2)\left( {1-\cos^{4}t(N)\over \sin^{2}t(N)} \right) +N+2\vphantom{{1-\cos^{4}t(N)\over \sin^{2}t(N)}}\right) +{N+2\over 4}\cos^{2}\xi
    \label{eqn:inter2}
\end{equation}
which obtains its maximum value
\begin{equation} {N^{2}-4\over 2}\cos^{2}t(N) + {N+2\over 2}(3+2\cos^{2}t(N)) \label{eqn:largescaleform}\end{equation}
at $\xi=\pi/2$.
This scaling form interpolates between Heisenberg scaling and the doubled SQL scaling. For $N+2=100$, Fig. \ref{fig:fig3} compares this scaling form to the maximal QFI on the path generated by $\vec{n}\cdot \vec{J}$ over all $\vec{n}$. One can see the two SQL scaling phases on the left and right that this functional form of the QFI does not describe. Note that the Heisenberg limit regime in a neighborhood of ${\pi\over 2}$ that occurs in the case of the bosonic one-axis twisting generator $J_{z}^{2}$ (i.e., all-to-all pairwise interactions of equal strength) does not exist for long, finite-range one-axis twisting even in the extreme case of $K=N/2$. In other words, full connectivity of interactions is necessary for achieving the Heisenberg limit.

\begin{figure}[t!]
\begin{center}
\includegraphics[scale=0.6]{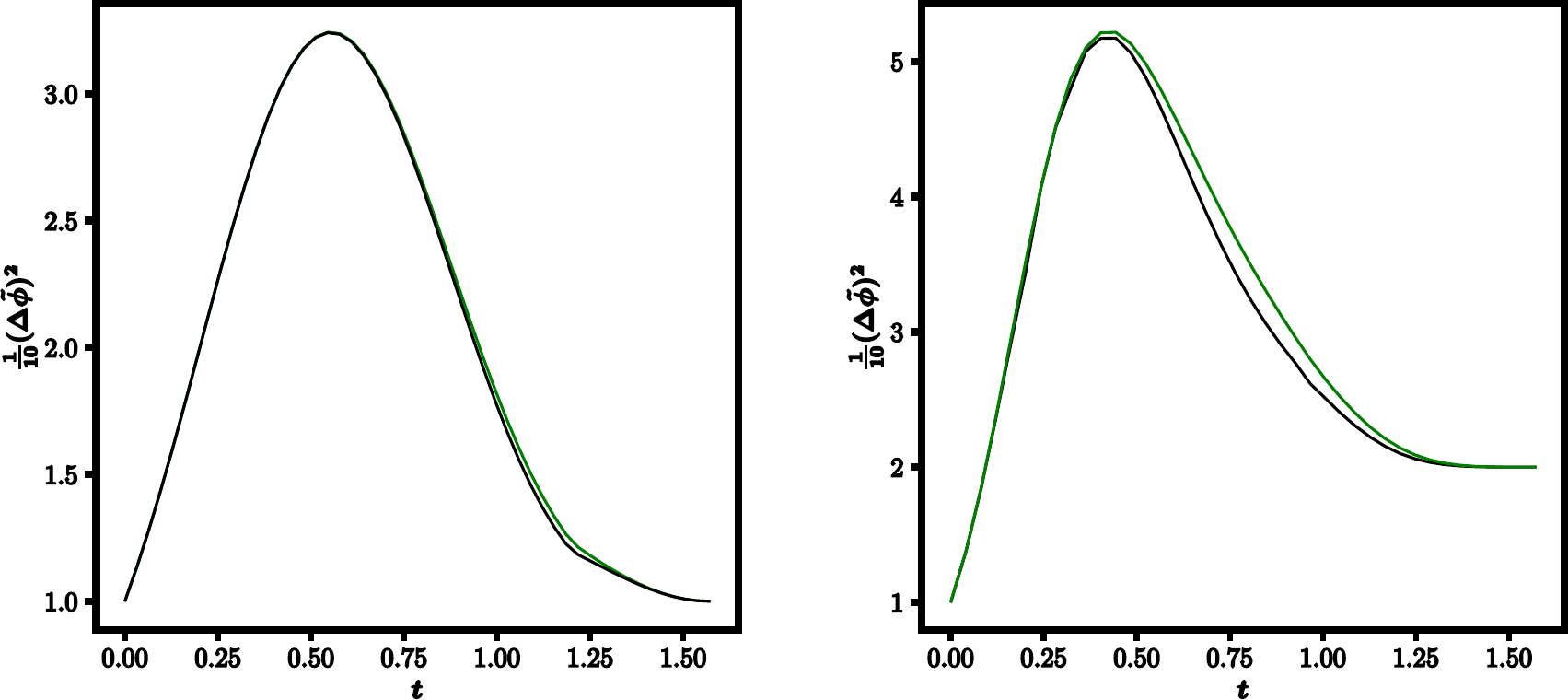}
 \caption{\label{fig:frmaxqfi}Comparison of maximal QFI for finite range, one-axis twisting (green) and maximal reciprocal of the method of moments error for a twist-untwist protocol with finite-range, one-axis twisting (black). Both values are multiplied by $N^{-1}$ so that the SQL corresponds to value 1. $N+2=10$, $K={N\over 4}=2$ (left); $N+2=10$, $K={N\over 2}=4$ (right). $(\Delta \tilde{\phi}^{2})_{\vec{n},\vec{m}}$ is optimized over $\vec{n}$ and $\vec{m}$ at $\phi=10^{-3}$ so that the computer program returns a finite result.
}
\end{center}
\end{figure}

\textbf{Heisenberg scaling.}
Because logarithmic corrections to $t=\Omega(N^{-1/2})$ scaling are covered by the previous regimes, it only remains to consider $t(N)={c\over \sqrt{N}}$. Plugging this scaling into (\ref{eqn:bigk}) gives QFI
\begin{equation}
    {N^{2}\sin^{2}\xi \over 2}(1+e^{-2c^{2}}\cos2\theta -2e^{-c^{2}}\cos^{2}\theta) + o(N^{2})
\end{equation}
which has a unique maximum ${N^{2}(1-e^{-2c^{2}})\over 2}$ at $\xi=\theta=\pi/2$. 

For both short and long finite-range one-axis twisting, the only range of interaction times that are relevant in the $K\rightarrow \infty$ with $K=\lambda N$ are captured by the Heisenberg-SQL scaling interpolations in (\ref{eqn:inter1}), (\ref{eqn:inter2}), respectively. Therefore, unlike  bosonic one-axis twisting which has only one asymptotic metrological phase described by Lemma \ref{lem:11}, finite-range one-axis twisting asymptotically has two metrological phases which are connected by the scaling forms in (\ref{eqn:inter1}), (\ref{eqn:inter2}) for the short range and long range cases, respectively.

\subsection{Finite-range twist-untwist protocols \label{sec:frtwun}}

In Ref.\cite{PhysRevResearch.4.013236}, the short finite-range ($K\le N/4$) twist-untwist protocol with $J_{y}$ rotation and $J_{y}$ readout was determined to asymptotically attain, at its optimal interaction time, 92\% of the one-axis twisting QFI in the $J_{y}$ direction at that same interaction time. That interaction time is in the Heisenberg scaling metrological phase defined by the interaction time scaling $t={c\over K^{1/2}}$. However, that work did not determine whether the best possible twist-untwist protocol (defined by optimal $\vec{n}$ and $\vec{m}$) can saturate the short finite-range one-axis twisting QFI on its optimal rotation path, which, as we showed in the previous subection, is indeed generated by $J_{y}$.  Ideally, to answer this question, one would use the analytical approach of Section \ref{sec:vvv} to identify the optimal $\vec{n},\vec{m}$ that allow a twist-untwist protocol to get as close as possible to the maximal value of (\ref{eqn:bestshort}) for this $t=O(K^{-1/2})$ metrological phase.

However, in the analysis of Section \ref{sec:vvv}, it was efficient to compute the $\phi \rightarrow 0$ limit of the empirical error $(\Delta \tilde{\phi})^{2}$ for the twist-untwist protocol defined by $J_{x}$ rotation and $J_{x}$ readout because when the one-axis twisting is generated by $J_{z}^{2}$, one can compute the small-$\phi$ expansions of the numerator and denominator of $(\Delta \tilde{\phi})^{2}$ by using the third and fourth moments of spin operators in SU(2) coherent states, which are easily derived. In contrast, the states that arise in the analogous calculation with a finite-range one-axis twisting generator $H_{K}$ are not restricted to the manifold of SU(2) coherent states, so calculating the third and fourth moments of the set $\lbrace J_{z},J_{+},J_{-} \rbrace$ for finite-range one-axis twist-untwist dynamics is considerably more complicated. We therefore provide numerical evidence in Fig. \ref{fig:frmaxqfi} for $N+2=10$, that for $K={N\over 4}$ in the short range case and $K={N\over 2}$ in the long range case that there is a finite-range twist-untwist protocol that saturates (differs by less than $N^{-1}$) the QFI in the Heisenberg scaling phase. In the case of short finite-range twist-untwist, the optimal protocol at the optimal interaction time on the time grid is given by rotation generator $\vec{n}\cdot \vec{J}$ with $\vec{n}=(0.86559,0.00020,0.50076)$ and measurement of $\vec{m}\cdot \vec{J}$ with $\vec{m}=(-0.99999,0.00033,-0.00009)$, whereas in the case of long finite-range twist-untwist, the optimal protocol at the optimal interaction time on the time grid is given by rotation generator $\vec{n}\cdot \vec{J}$ with $\vec{n}=(0.00010,0.92735,0.37419)$ and measurement of $\vec{m}\cdot \vec{J}$ with $\vec{m}=(-0.99999,0.00044,-0.00009)$. In both cases, the vectors $\vec{n}$ and $\vec{m}$ were optimized over the hemisphere $\theta \in (0,\pi)$, $\varphi \in (0,\pi)$.

\section{Conclusion}
In this work we have shown that twist untwist protocols with total spin readout are asymptotically sufficient for estimation of SU(2) rotations applied to one-axis twisted probe states for almost all interaction times, and provided numerical evidence that the same statement holds when the one-axis twisting generator is replaced by finite-range one-axis twisting of a lattice spin-1/2 system. Other generators of spin squeezing such as the two-axis countertwisting interaction \cite{PhysRevA.47.5138} and the twist-and-turn interaction \cite{PhysRevA.92.023603} are known to form metrologically useful probe states for interferometry. Therefore, one direction of future research is to fully classify the sufficiency of the method of moments error of optimal twist-untwist protocols for saturating the QCRI for interferometry using probe states generated by Hamiltonians at most quadratic in the $J_{x}$, $J_{y}$, $J_{z}$ operators \cite{PhysRevA.94.042327}. 
Finally, we note that the QFI formulas appearing in \ref{sec:qfifr} that allow to analyze the metrological phase diagram for finite-range one-axis twisting in Section \ref{sec:foat} are derived using a diagrammatic method that keeps track of all contributions to the relevant inner products. Design of analogous methods for two and higher dimensions will lead to a rigorous analysis of the interplay of atom number $N$, interaction range $K$, and dimension $D$ in the QFI for finite-range one-axis twisting.

\acknowledgements
The authors were supported by the Laboratory Directed Research and Development (LDRD) program of Los Alamos National Laboratory (LANL) under project number 20210116DR. Michael J. Martin was also supported by supported by the U.S. Department of Energy, Office of Science, National Quantum Information Science Research Centers, Quantum Science Center. Los Alamos National Laboratory is managed by Triad National Security, LLC, for the National Nuclear Security Administration of the U.S. Department of Energy under Contract No. 89233218CNA000001.

\onecolumngrid
\bibliography{phasebib.bib}

\onecolumngrid
\appendix
\section{\label{sec:app1}Calculation of (\ref{eqn:rfrf}) }
The first line of (\ref{eqn:rfrf}) is a standard computation (see second moments of spin operators in SU(2) coherent states \cite{wangmolm,PhysRevA.68.012101}) and yields
\begin{equation}
    -N\phi \text{Re}\langle \zeta=1 \vert e^{itJ_{z}^{2}}J_{x}^{2}e^{-itJ_{z}^{2}}\vert \zeta=1\rangle=-N\phi \left( {N^{2}+N\over 8}+{N(N-1)\over 8}\cos^{N-2}2t \right).
\end{equation}
The second line of (\ref{eqn:rfrf}) is given by
\begin{eqnarray}
    &{}& 2\phi \text{Re}\langle\zeta=1\vert e^{itJ_{z}^{2}}J_{x}e^{-itJ_{z}^{2}}J_{x}e^{itJ_{z}^{2}}J_{x}e^{-itJ_{z}^{2}}\vert \zeta=1\rangle = \nonumber \\
    &{}& {\phi \over 4}\text{Re} \langle \zeta=1\vert \left( e^{2it(J_{z}-{1\over 2})}J_{+}+e^{-2it(J_{z}+{1\over 2})} J_{-}\right) \left( J_{+}+J_{-} \right) \nonumber \\
    &{}& \left( J_{+}e^{2it(J_{z}+{1\over 2})}+J_{-}e^{-2it(J_{z}-{1\over 2})}  \right) \vert \zeta=1\rangle  \\
    &{}& = {\phi \over 4}\text{Re} \left[ \vphantom{\sum_{0}^{\infty}} \langle \zeta =e^{2it} \vert  J_{+}^{3} +J_{+}J_{-}J_{+} \vert \zeta=e^{-2it}\rangle  +c.c.  \right. \nonumber \\ 
    &{}& \left. +  \langle \zeta =e^{2it} \vert  J_{+}^{2}J_{-} \vert \zeta =e^{2it}\rangle + c.c. \right. \nonumber \\
    &{}& \left. +  \langle \zeta =e^{-2it} \vert  J_{-}J_{+}^{2} \vert \zeta =e^{-2it}\rangle + c.c. \vphantom{\sum_{0}^{\infty}} \right] \nonumber \\
    &=& {\phi \over 4} \left[ \vphantom{\sum_{0}^{\infty}} {N(N-1)(N-2)\over 2} \cos^{N-3}2t + {N^{2}} \cos^{N-1}2t  + 2N(N-1)\cos 2t \right. \nonumber \\
    &{}& \left. + {N(N-1)(N-2)\over 2}\cos 2t   \vphantom{\sum_{0}^{\infty}} \right].\label{eqn:aaaa}
\end{eqnarray}
So the full result for (\ref{eqn:rfrf}) is 
\begin{equation}
    \del_{\phi}\langle J_{x}\rangle_{\ket{\psi_{\phi}((1,0,0))}} =-{N\phi \over 8}\left( N^{2}+N + N(N-1)\cos^{N-2}2t \right)+ (\ref{eqn:aaaa}) + O(\phi^{3})
    \label{eqn:resres}.
\end{equation}

Taking $t={c\over N^{\alpha}}$, $0<\alpha<1/2$ gives 
\begin{eqnarray}
    \left( \del_{\phi}\langle J_{x}\rangle_{\ket{\psi_{\phi}((1,0,0))}} \right)^{2}&\sim& \left( \left( {N^{3}\over 8}+{N^{2}\over 8} \right)(1-\cos2t) - {N\over 4}\cos 2t \right)^{2} \nonumber \\
    &\sim& \left( N^{3}\sin^{2}{c\over N^{\alpha}}+N^{2}\sin^{2}{c\over N^{\alpha}}+N\cos {2c\over N^{\alpha}} \right)^{2} {\phi^{2}\over 16}
\end{eqnarray}

\section{\label{sec:app2}Calculation of $J_{\pm}$ moments  for (\ref{eqn:varzro}) }

Using the results of \ref{sec:app1}, (\ref{eqn:ghgh}) is simplified to
\begin{eqnarray}
     \phi^{-2}\text{Var}_{\ket{\psi_{\phi}((1,0,0))}}J_{x}&=&{N^{4}\over 32}+{N^{3}\over 32} + \left( {N^{4}\over 16}+{N^{3}\over 16}-{N^{2}\over 8} \right)\cos 2t \nonumber \\
     &{}& + \langle e^{itJ_{z}^{2}}J_{x}e^{-itJ_{z}^{2}}J_{x}^{2}e^{itJ_{z}^{2}}J_{x}e^{-itJ_{z}^{2}}\rangle_{\ket{\zeta=1}}\nonumber \\
     &{}&+O(\phi^{2}).
     \label{eqn:ghgh2}
 \end{eqnarray}
 Since for $t={c\over N^{\alpha}}$, $0<\alpha<1/2$, it follows that $\cos^{N}t \rightarrow 0$, one finds that the second line of (\ref{eqn:ghgh2}) has the asymptotic form
 \begin{eqnarray}
     \langle e^{itJ_{z}^{2}}J_{x}e^{-itJ_{z}^{2}}J_{x}^{2}e^{itJ_{z}^{2}}J_{x}e^{-itJ_{z}^{2}}\rangle_{\ket{\zeta=1}}&\sim &{1\over 16}\langle J_{+}^{3}J_{-} + J_{+}J_{-}J_{+}J_{-}+J_{+}^{2}J_{-}^{2}+J_{+}J_{-}^{3} \rangle_{\ket{\zeta=e^{2it}}} \nonumber \\
     &+& {1\over 16}\langle J_{-}J_{+}^{3} +J_{-}^{2}J_{+}^{2}+ J_{-}J_{+}J_{-}J_{+}+J_{-}^{3}J_{+} \rangle_{\ket{\zeta=e^{-2it}}} \nonumber \\
     &\sim &{N^{4}\over 64}\left(1+\cos 4t\right).
 \end{eqnarray}
 Substituting this result into (\ref{eqn:ghgh2}) and taking $t={c\over N^{\alpha}}$, $0<\alpha<1/2$ gives
 \begin{equation}
     \text{Var}_{\ket{\psi_{\phi}((1,0,0))}}J_{x} \sim {N^{4}\phi^{2}\over 8}\sin^{4}{c\over N^{\alpha}}
 \end{equation}
 which gives (\ref{eqn:varzro}) for large $N$.
 
 \section{QFI for finite-range one-axis twisting\label{sec:qfifr}}
  $N+2$ atoms on a ring, $N$ even. We use 
  \begin{eqnarray}
     &{}& \text{Var}_{e^{-itH_{K}}\ket{+}^{\otimes N}}\left( \sin \xi \cos \theta J_{x}+ \sin \xi \sin\theta J_{y} + \cos\xi J_{z} \right) = \nonumber \\
      &{}& {\sin^{2}\xi\over 2}\left( \text{Re}\left( e^{2i\theta} \langle J_{-}^{2}\rangle\right) + \langle J_{-}J_{+}\rangle \right) \nonumber \\
      &{}& +{\sin 2\xi\over 2}\left( \text{Re}\left( e^{i\theta}\langle J_{z}J_{-}\rangle\right) +\text{Re}\left( e^{-i\theta}\langle J_{z}J_{+}\rangle\right) \right) + \cos^{2}\xi\langle J_{z}^{2}\rangle \nonumber \\
      &{}& -\sin^{2}\xi\cos^{2}\theta \left( \text{Re}\langle J_{+}\rangle \right)^{2}
  \end{eqnarray}
  and a table of first and second moments of the $\mathfrak{su}(2)$ elements $\lbrace J_{+},J_{-},J_{z}\rbrace$. Note that $\langle J_{z} \rangle_{e^{-itH_{K}}\ket{+}^{\otimes N}} =0$ and $\langle J_{y} \rangle_{e^{-itH_{K}}\ket{+}^{\otimes N}} =0$.

  \subsection{$1\le K \le {N\over 4}$}
  
  \begin{small}
   \begin{eqnarray}
     &{}&\text{Var}_{e^{-itH_{K}}\ket{+}^{\otimes N}}\left( \sin \xi \cos \theta J_{x}+ \sin \xi \sin\theta J_{y} + \cos\xi J_{z} \right)=\nonumber \\ 
     &{}& {\sin^{2}\xi\over 2} \left( {N+2\over 2} +{N+2\over 4}(N+1-4K)\cos^{4K}t + {N+2\over 2} {\cos^{2K}t-\cos^{4K}t \over \sin^{2}t} \right. \nonumber \\
     &{}&\left. +{N+2\over 2}\cot^{2}t(1-\cos^{2K}t) \right)\nonumber \\
      &{}& +{\sin^{2}\xi\cos2\theta\over 2} \left( {N+2\over 4}(N+1-4K)\cos^{4K}t  +{N+2\over 2}\cos^{4K}t{ \left({\cos 2t\over \cos^{2}t}\right)^{K+1}-{\cos 2t\over \cos^{2}t}\over {\cos 2t\over \cos^{2}t}-1} \right. \nonumber \\
      &{}& \left. +{N+2\over 2}{\cos^{2K}t\cos^{K-1}2t \left( \left( {\cos 2t\over \cos^{2}t}\right)^{K}-1\right) \over {\cos 2t\over \cos^{2}t} -1} \right)\nonumber \\
      &{}& +{\sin 2\xi \over 2}(N+2)K\sin t\sin\theta\cos^{2K-1}t +{(N+2) \cos^{2}\xi \over 4}- {\sin^{2}\xi \cos^{2}\theta (N+2)^{2} \over 4}\cos^{4K}t \nonumber \\
     &=&  {\sin^{2}\xi\over 2} \left( {N+2\over 2} -(1+4K)\cos^{4K}t + {N+2\over 2} {\cos^{2K}t-\cos^{4K}t \over \sin^{2}t}+{N+2\over 2}\cot^{2}t(1-\cos^{2K}t) \right) \nonumber \\
     &{}& +{\sin^{2}\xi\cos2\theta\over 2} \left( -(1+4K)\cos^{4K}t  +{N+2\over 2}\cos^{4K}t{ \left({\cos 2t\over \cos^{2}t}\right)^{K+1}-{\cos 2t\over \cos^{2}t}\over {\cos 2t\over \cos^{2}t}-1} \right. \nonumber \\
      &{}& \left. +{N+2\over 2}{\cos^{2K}t\cos^{K-1}2t \left( \left( {\cos 2t\over \cos^{2}t}\right)^{K}-1\right) \over {\cos 2t\over \cos^{2}t} -1} \right)\nonumber \\
      &{}& +{\sin 2\xi \over 2}(N+2)K\sin t\sin\theta\cos^{2K-1}t +{(N+2) \cos^{2}\xi \over 4}
     \label{eqn:smallk}
      \end{eqnarray}
  \end{small}

 \subsection{${N/4}<K\le {N\over 2}$}

 \begin{eqnarray}
     &{}&\text{Var}_{e^{-itH_{K}}\ket{+}^{\otimes N}}\left( \sin \xi \cos \theta J_{x}+ \sin \xi \sin\theta J_{y} + \cos\xi J_{z} \right)= \nonumber \\
     &{}& {\sin^{2}\xi\over 2}\left(  {N+2\over 2} \left( {1-\cos^{2(N+2-2K)}t\over\sin^{2}t}\right) + {(N+2)(3K-N-1)\over 2}\cos^{2(N+1-2K)}t \right. \nonumber \\
     &{}& \left. + {N+2\over 4}(N+1-2K)\cos^{2(N-2K)}t \right)\nonumber \\
     &{}& + {\sin^{2}\xi\cos2\theta \over 2} \left( {N+2\over 2}\cos^{2K-1}2t { \left( {\cos^{2}t\over \cos2t} \right)^{N+2-2K} - {\cos^{2}t\over \cos2t}\over {\cos^{2}t\over \cos2t}-1} \right. \nonumber \\
     &{}& \left. +{(N+2)(3K-N-1)\over 2}\cos^{2(N+1-2K)}t\cos^{4K-N-2}2t \right. \nonumber \\
     &{}& \left. +{(N+2)(N+1-2K)\over 4}\cos^{2(N-2K)}t\cos^{4K-N}2t \vphantom{ \left( {\cos^{2}t\over \cos2t} \right)^{N+2-2K} - {\cos^{2}t\over \cos2t}\over {\cos^{2}t\over \cos2t}-1}\right) \nonumber \\
     &{}& +{\sin 2\xi \over 2}(N+2)K\sin t\sin\theta\cos^{2K-1}t +{(N+2) \cos^{2}\xi \over 4} \nonumber \\
     &{}& - {\sin^{2}\xi \cos^{2}\theta (N+2)^{2} \over 4}\cos^{4K}t
     \label{eqn:bigk}
 \end{eqnarray}

\end{document}